\documentclass[12pt]{article}
\usepackage{graphicx} 
\usepackage[sorting=none]{biblatex}
\usepackage{authblk}
\usepackage{hyperref}
\usepackage{amsmath, amssymb}
\usepackage{biblatex2bibitem}
\usepackage{mathtools}
\title{\textbf{SVD Entanglement Entropy of Chiral Dirac Oscillators}}
\author[1,2]{Yuvraj Singh \thanks{ys987@snu.edu.in}}
\author[1]{Rabin Banerjee \thanks{rabin@bose.res.in}}
\affil[1]{ S.N. Bose National Centre for Basic Sciences, JD Block, Sector III, Salt Lake, Kolkata 700106, India }
\affil[2]{Department of Physics, Shiv Nadar Institution of Eminence, Gautam Buddha Nagar, Uttar Pradesh 201314, India }
\date{July 2024}
\begin{document}
\maketitle
\begin{abstract}
    We discuss the SVD entanglement entropy, which has recently come up as a successor to the pseudo entropy. This paper is a first-of-its-kind application of SVD entanglement entropy to a system of chiral Dirac oscillators which prove to be a natural system to study the SVD formalism because the two chiral oscillator ground states can be taken as the pre-selected and post-selected states. We argue how this alternative for entanglement entropy is better and more intuitive than the von Neumann one to study quantum phase transition. It is shown that SVD entropy diverges at the critical point, matching von Neumann entropy in the left-handed regime but differing in the right-handed regime. We also provide as an illustrative example, a new generalized proof of the SVD entanglement entropy being $\log2$ for a pair of Bell states that differ from each other by relative phases.
\end{abstract}

\textbf{INTRODUCTION}\\
Entanglement entropy is a measure of the quantum correlation or entanglement between two subsystems in a quantum system. It has become an important parameter in studying the quantum properties of any emergent quantum system. Entanglement entropy is usually computed by the definition of entropy given by von Neumann \cite{Nielsen_Chuang_2010} and is called the von Neumann entanglement entropy. Here, we consider a single state and compute the entanglement entropy which is interpreted as a measure of quantum correlations among the degrees of freedom for the given state. Recently however a new way has been devised to calculate entanglement entropy using the SVD (Singular Value Decomposition) entanglement entropy \cite{svd}. In this formalism, we take two states and form a density matrix from these states rather than taking the same state. These states are called pre-selection and post-selection states. Then the entanglement entropy gives a measure of quantum correlations among the degrees of freedom for the post-selection and pre-selection states. It is observed that this result of SVD entanglement entropy comes out close to the usual von Neumann entanglement entropy \cite{svd}. For further developments and generalizations refer to \cite{he2023pseudo,narayan2024further, shinmyo2024pseudo,guo2024sum,he2024entanglement,kanda2024entanglement,grieninger2024temporal,anegawa2024black,nandy2024probing,guo2024relation,guo2024pseudoentropy,nakata2021new,narayan2303notes,doi2023pseudoentropy}.

Here we concern ourselves with the entanglement entropy related to the states of the Chiral Dirac oscillator which is a relativistic fermionic harmonic oscillator in the presence of a constant magnetic field. These oscillators were first studied in \cite{ito, M} and prove to be a very interesting system to study in quantum optics and condensed matter physics. In recent decades a lot of work has been put into the theoretical \cite{Ivanov_2013,PhysRevA.90.042111} as well as experimental advancement \cite{PhysRevLett.111.170405} of this in terms of its thermal properties \cite{Okon2021}, further generalizations \cite{10.1063/1.4905085,PhysRevD.91.045032} and its applications to cosmology and gravitational systems \cite{Oliveira2024, Andrade2014,Bakke2013} have also come up. Non-relativistic oscillators and chirality have been studied in a fundamental and field-theoretic manner by \cite{ banerjee2000canonical,banerjee1998chiral, banerjee2012dual} its connection to material science and topological Chern-Simons gauge theory has been pointed out in \cite{10.1063/5.0214919, PhysRevD.41.661}. In \cite{chiral}, it was shown that these Dirac oscillators in the presence of a constant magnetic field exhibit a quantum phase transition. This study calculated the analytical expression for order parameter, and canonical quantum fluctuations for the quantum phase transition, furthermore, entanglement properties between the degrees of freedom in the relativistic ground state were also studied. The entanglement entropy turns out to be divergent at the point of phase transition which is seen as a common feature of quantum phase transition \cite{landau2005guide,pethick2008bose,stanley1987introduction,sachdev2011quantum}.\par

In this work, we have shown how SVD entanglement entropy can be used to calculate the entanglement entropy for a system exhibiting chiral behavior. Our results show that the behavior of SVD entanglement entropy is similar to \cite{chiral} but the physical interpretation is more intuitive than the von Neumann entanglement entropy. In both cases, the value of entanglement entropy is divergent at the critical point indicative of a quantum phase transition \cite{svd,sachdev2011quantum}.
\par
As a prelude to the study of chiral Dirac oscillators, we provide an illustrative example, a new generalized proof of the SVD entanglement entropy being $\log2$ for a pair of Bell states that differ from each other by relative phases. This is a generalization of a result given in \cite{svd} where they have considered a specific example of a Bell state. 
\par
The organization of this paper is as follows. In the first section, we introduce the SVD entanglement entropy and its basic properties along with the generalized example for Bell states. In the second section, we introduce the formalism related to the chiral Dirac oscillator.  In the third section, we introduce our main result and compare it with the old result for entanglement entropy. In the last section, we conclude our essay.

\section{SVD ENTANGLEMENT ENTROPY}
There have been numerous ways that try to incorporate a post-selection and pre-selection state into the calculation of entanglement entropy, as it is a useful formalism for treating problems in quantum gravity, black hole information paradox, condensed matter physics \cite{horowitz2004black,gottesman2004comment,nakata2021new,salek2014negative,fullwood2023dynamical,tu2022renyi, zhu2011quantum,ferrie2014single,combes2014quantum} and quantum information theory \cite{aharov1988spin, ramos2020measurement, arvidsson2020quantum, lupu2022negative}. 
The aim of this section is to introduce this new definition of entropy as elaborated in \cite{svd}.
We are already familiar with the von Neumann entropy for a bipartite system.
If $\rho$ represents the density operator of a system then, the von Neumann entropy is given by \cite{Nielsen_Chuang_2010},
\[
S = -Tr(\rho \log\rho)
\]
But this definition of entropy uses only a single state and that state may be entangled or not. 
\par
The SVD (singular value decomposition) entanglement entropy uses two states a pre-selected state and a post-selected state to calculate the entanglement entropy. It is proposed to be a natural generalization of the von Neumann entanglement entropy. \par Consider a Hilbert space that is decomposed into two Hilbert spaces $\mathcal{H} = \mathcal{H}_A \otimes \mathcal{H}_B$ if we take $|\psi_1 \rangle$ as the pre-selected state and $|\psi_2\rangle$ as the post-selected state, then we can define the reduced transition matrix\cite{svd},
\begin{equation}
    \tau^{1|2} := \frac{|\psi_1\rangle\langle\psi_2|}{\langle\psi_1|\psi_2\rangle}   
\end{equation}

whereas the unnormalized version is,
\[
\tau^{1|2} = |\psi_1\rangle\langle\psi_2|
\]
Now if we take the partial trace,
\[
\tau^{1|2}_A  = Tr_B[\tau^{1|2}]
\]
Using this we define,
\begin{equation}
    \rho^{1|2}_A := \frac{\sqrt{(\tau^{1|2}_A)^{\dagger}(\tau^{1|2}_A)}}{Tr\left[\sqrt{(\tau^{1|2}_A)^{\dagger}(\tau^{1|2}_A)}\right]}   
\end{equation}
This operator is a density operator. Then SVD entanglement entropy is,
\begin{equation}
    S_{SVD}(\tau^{1|2}_A) := - Tr[\rho^{1|2}_A \log\rho^{1|2}_A]    
\end{equation}
Hence, the functional forms of SVD and von Neumann entanglement entropy are similar.\\
Now, we try to summarize the general definition and physical interpretation (in the context of information theory).\\
\textbf{General Definition: }
Given a space $\mathbb{M}_m$ of $m \times m$ matrices. Then $S_{SVD}$ is defined by,
\[
S_{SVD}: \mathbb{M}_m \longrightarrow \left[ 0,\infty\right)
\]
\[
S_{SVD}(A) = S_{vN}\left(\frac{|A|}{||A||_1}\right) = -\sum^m_{i=1} \frac{\sigma_i}{||A||_1}\log \left(\frac{\sigma_i}{||A||_1}\right)
\]
where $|A| := \sqrt{A^{\dagger}A}$ and, $||A||_1 := Tr|A| = \sum^{m}_{i=1}\sigma_i$ ($\sigma_i$ are eigenvalues of $|A|$). Also note if $A=0$ then $S_{SVD} = 0$.\\
\textbf{Physical Interpretation: }
von Neumann entropy coincides with the average number of Bell pairs that can be distilled from a given bipartite state via local operations and classical communication\cite{PhysRevA.53.2046,Nielsen_Chuang_2010}.
This result can also be proved for SVD entropy.\cite{svd}\\
We can also diagonalize the reduced transition matrix,
\[
\tau^{1|2} = U\cdot\Lambda \cdot V^\dagger
\]
where $U$ and $V$ are unitary matrices, and $\Lambda$ is the diagonal matrix such that $\Lambda = diag(\lambda_1,\ldots,\lambda_{d_A})$ and $\lambda_i \geq 0,  \forall i \in \{1,\ldots,d_A\}$. Then we can define the normalized eigenvalues as:
\[
\hat{\lambda}_i = \frac{\lambda_i}{\sum\lambda_i}
\]
Clearly, $\hat{\lambda}_i$ are eigenvalues of $\rho^{1|2}_A$, then we have:
\[
S_{SVD} = -\sum^{d_A}_{i} \hat{\lambda}_i \log\hat{\lambda}_i
\]

\subsection{\centering{General Result for the Bell States}}
Now, we make a general statement about the SVD entropy of the Bell states. In \cite{svd} two states belonging to a bipartite system of bell states ($\mathcal{H}_A \otimes \mathcal{H}_B$) have been considered:
\begin{equation}
   \begin{split}
       |\psi_1\rangle &= \frac{1}{\sqrt{2}}(|00\rangle + |11\rangle)\\
       |\psi_2\rangle &= \frac{1}{\sqrt{2}}(|00\rangle + \phi|11\rangle)
   \end{split} 
   \label{svd}
\end{equation}
where $\phi \in \mathbf{C}$, $|\phi|=1$, and $|\psi_1\rangle$ and $|\psi_2\rangle$ are the pre-selected and post-selected states respectively.\\
Now we consider the following generalized states belonging to a bipartite system of bell states ($\mathcal{H}_A \otimes \mathcal{H}_B$). 
\begin{equation}
\begin{split}
    |\psi_1\rangle &= a|\alpha\beta\rangle + b|\gamma\delta\rangle,\\
    |\psi_2\rangle &= c|\tilde{\alpha}\tilde{\beta}\rangle + d|\tilde{\gamma}\tilde{\delta}\rangle 
\end{split}
\end{equation}
where $\alpha, \beta, \gamma, \delta, \tilde{\alpha}, \tilde{\beta}, \tilde{\gamma}$, and $\tilde{\delta} \in \{0,1\}$ and $a,b,c,d$ $\in \mathbf{C}$, where $|a|, |b|, |c|, |d| = 1$.
The transition matrix is given by,
\[
\tau^{1|2} = \frac{|\psi_1\rangle\langle\psi_2|}{\langle\psi_1|\psi_2\rangle}
\]
\[
\tau^{1|2} = \frac{ac^*|\alpha\beta\rangle\langle\Tilde{\alpha}\Tilde{\beta}| + ad^*|\alpha\beta\rangle\langle\Tilde{\gamma}\Tilde{\delta}| + ba^*|\gamma\delta\rangle\langle\Tilde{\alpha}\Tilde{\beta}| + bd^{*}|\gamma\delta\rangle\langle\Tilde{\gamma}\Tilde{\delta}|}{a^*c\langle\alpha\beta|\Tilde{\alpha}\Tilde{\beta}\rangle + a^{*}d\langle\alpha\beta|\Tilde{\gamma}\Tilde{\delta}\rangle + b^*c\langle\gamma\delta|\Tilde{\alpha}\Tilde{\beta}\rangle + b^{*}d\langle\gamma\delta|\Tilde{\gamma}\Tilde{\delta}\rangle}
\]
The reduced transition matrix is,
\[
\tau^{1|2}_A = Tr_B(\tau^{1|2})
\]
\[
\tau^{1|2}_A = \frac{ac^*|\alpha\rangle\langle\Tilde{\alpha}|(\langle\beta|\Tilde{\beta}\rangle) + ad^{*}|\alpha\rangle\langle\Tilde{\gamma}|(\langle\beta|\Tilde{\delta}\rangle) + bc^*|\gamma\rangle\langle\Tilde{\alpha}|(\langle\delta|\Tilde{\beta}\rangle) + bd^{*}|\gamma\rangle\langle\Tilde{\gamma}|(\langle\delta|\Tilde{\delta}\rangle)}{a^*c\langle\alpha\beta|\Tilde{\alpha}\Tilde{\beta}\rangle + a^*d\langle\alpha\beta|\Tilde{\gamma}\Tilde{\delta}\rangle + b^*c\langle\gamma\delta|\Tilde{\alpha}\Tilde{\beta}\rangle + b^*d\langle\gamma\delta|\Tilde{\gamma}\Tilde{\delta}\rangle}
\]
Since these are Bell states:
\begin{itemize}
    \item \textit{\underline{Case I:}} If without loss of generality we assume that $\alpha=\beta=0$ $\implies$ $\gamma$, $\delta = 1$ (Bell states). 
    \[
    \tau^{1|2}_A =\frac{ac^*|0\rangle\langle\Tilde{\alpha}|(\langle0|\Tilde{\beta}\rangle) + ad^{*}|0\rangle\langle\Tilde{\gamma}|(\langle0|\Tilde{\delta}\rangle) + bc^*|1\rangle\langle\Tilde{\alpha}|(\langle1|\Tilde{\beta}\rangle) + bd^{*}|1\rangle\langle\Tilde{\gamma}|(\langle1|\Tilde{\delta}\rangle)}{a^*c\langle00|\Tilde{\alpha}\Tilde{\beta}\rangle + a^*d\langle00|\Tilde{\gamma}\Tilde{\delta}\rangle + b^*c\langle11|\Tilde{\alpha}\Tilde{\beta}\rangle + b^*d\langle11|\Tilde{\gamma}\Tilde{\delta}\rangle}
    \]
    Also for the denominator to be non-zero $\Tilde{\alpha}=\Tilde{\beta}$ and $\Tilde{\gamma}=\Tilde{\delta}$. Since we are working with Bell states $\implies$ $\Tilde{\alpha} \neq \Tilde{\gamma}$.\\
    Again, without loss of generality assume that, $\Tilde{\gamma} = 0$ and $\Tilde{\alpha} = 1$, then,
    \begin{equation}
        \tau^{1|2}_A = \frac{\left[ ad^*|0\rangle\langle0| + bc^{*}|1\rangle\langle1|\right]}{(a^*d+b^*c)}
    \end{equation}
    If we had used the complimentary values i.e. $\tilde{\gamma} =1$ and $\tilde{\alpha}=0$, then,
    \begin{equation}
        \tau^{1|2}_A = \frac{\left[ ac^*|0\rangle\langle0| + bd^{*}|1\rangle\langle1|\right]}{(a^*c+b^*d)}
    \end{equation}

    \item \textit{\underline{Case II:}} If $\alpha=\delta$ then $\beta=\gamma$, $\alpha \neq \beta$ and $\gamma \neq \delta$, so without loss of generality if we assume that $\alpha=0$ then $\beta = 1$, $\gamma=1$, and $\delta=0$. 

    \[
    \tau^{1|2}_A =\frac{ac^*|0\rangle\langle\Tilde{\alpha}|(\langle1|\Tilde{\beta}\rangle) + ad^{*}|0\rangle\langle\Tilde{\gamma}|(\langle1|\Tilde{\delta}\rangle) + bc^*|1\rangle\langle\Tilde{\alpha}|(\langle0|\Tilde{\beta}\rangle) + bd^{*}|1\rangle\langle\Tilde{\gamma}|(\langle0|\Tilde{\delta}\rangle)}{a^*c\langle01|\Tilde{\alpha}\Tilde{\beta}\rangle + a^*d\langle01|\Tilde{\gamma}\Tilde{\delta}\rangle + b^*c\langle10|\Tilde{\alpha}\Tilde{\beta}\rangle + b^*d\langle10|\Tilde{\gamma}\Tilde{\delta}\rangle}
    \]
    Also for the denominator to be non-zero $\Tilde{\alpha} \neq \Tilde{\beta}$ and $\Tilde{\gamma} \neq \Tilde{\delta}$. Since we are working with Bell states $\implies$ $\Tilde{\alpha} = \Tilde{\delta}$ and $\Tilde{\beta} = \Tilde{\gamma}$.\\
    Again, without loss of generality assume that, $\Tilde{\alpha}=1$ and $\Tilde{\beta}=0$, then,
    \begin{equation}
        \tau^{1|2}_A = \frac{\left[ ad^{*}|0\rangle\langle0| + bc^*|1\rangle\langle1|\right]}{(a^*d+b^*c)}
    \end{equation}
    If we had assumed the complimentary values i.e. $\tilde{\alpha}=0$ and $\tilde{\beta}=0$, then,
    \begin{equation}
        \tau^{1|2}_A = \frac{\left[ ac^{*}|0\rangle\langle0| + bd^*|1\rangle\langle1|\right]}{(a^*c+b^*d)}
    \end{equation}
\end{itemize}
\par
Now, we evaluate the density matrix from the reduced transition matrix. 
\[
    \rho^{1|2}_A = \frac{\sqrt{(\tau^{1|2}_A)^\dagger \tau^{1|2}_A}}{Tr\left(\sqrt{(\tau^{1|2}_A)^\dagger \tau^{1|2}_A}\right)}
\]
\[
    \rho^{1|2}_A = \frac{\sqrt{\frac{|ad|^2|0\rangle\langle0| + |bc|^2|1\rangle\langle1|}{(a^*d +bc^*)(ad^* +b^*c)}}}{Tr\left( \sqrt{\frac{|ad|^2|0\rangle\langle0| + |bc|^2|1\rangle\langle1|}{(a^*d +bc^*)(ad^* +b^*c)}}\right)}
\]
\[
\rho^{1|2}_A = \frac{|0\rangle\langle0|+|1\rangle\langle1|}{2}
\]
\begin{equation}
    S_{SVD} = -Tr\left(\rho^{1|2}_A \log\rho^{1|2}_A \right) = \log2
\end{equation}
Thus the entropy for the maximally entangled Bell states is $\log2$ which is equal to the value of von Neumann entropy for Bell states. This result confirms the validity of SVD entanglement entropy as a generalization of the usual von Neumann entanglement entropy for two different states.

\section{CHIRAL DIRAC OSCILLATORS}
In the previous example, introducing \textit{by hand} a phase among the Bell states, the pre-selected and post-selected states were defined. Chiral Dirac oscillators, on the other hand, provide a natural setting for obtaining these states by considering the opposite chiralities. 
The Dirac equation gives the Dirac oscillator in its usual framework \cite{ito, M}. The Dirac oscillator modifies the usual Dirac equation by introducing a linear potential term. Specifically, the momentum operator $\textbf{p}$ in the Dirac equation is replaced with $\textbf{p} - \iota m \beta \omega \textbf{r}$ and we get the Hamiltonian as:
\begin{equation}
\iota\hbar \frac{\partial |\psi\rangle}{\partial t} = [c \boldsymbol{\alpha} (\textbf{p} -\iota m\beta\omega\textbf{r} ) + \beta m c^2]|\psi\rangle
\label{1}  
\end{equation}
where $m$ is the relativistic mass of the fermion, $-e$ is the electric charge, $|\psi\rangle$ stands for the Dirac four-component spinor, \textbf{r} and \textbf{p} represent position and momentum operators, $\omega$ is the Dirac oscillator frequency, $c$ is the speed of light and $\beta := diag(\mathbf{I}, - \mathbf{I})$, $\alpha_j:= off-diag(\sigma_j,\sigma_j)$ are the Dirac matrices related to the usual Pauli matrices ($\sigma_1, \sigma_2, \sigma_3$). This relativistic oscillator involves two phonon modes. Now, if we introduce the magnetic field by minimal coupling $\textbf{p} \longrightarrow \textbf{p} + e \textbf{A}$ where \textbf{A} is the vector potential related to the magnetic field. In the two-dimensional setup, the Dirac matrices become the Pauli matrices \cite{chiral,banerjee2020non}, then \eqref{1} becomes:
\begin{equation}
    \iota \hbar \frac{\partial |\psi\rangle}{\partial t} = \left[ c \sum ^2_{j=1} \sigma_j(p_j - \iota m\beta\sigma_z x_j + eA_j) + \sigma_z m c^2\right]|\psi\rangle
    \label{2}
\end{equation}
where $|\psi \rangle$ is a two-component spinor that mixes spin-up and spin-down with positive and negative energies. The non-minimal coupling of Dirac oscillator and the minimal coupling of the magnetic field endows the particle with intrinsic left-handed and right-handed chirality respectively \cite{chiral, PhysRevLett.99.123602, PhysRevA.76.041801}. This chiral distribution of the oscillator terms into right-handed and left-handed chirality can be probed by using the Jaynes-Cummings (JC) and Anti-Jaynes-Cummings (AJC) Hamiltonian respectively which have been extensively studied \cite{JC} before. The Jaynes-Cummings model is a cornerstone of quantum optics, describing the interaction between a two-level atom (e.g., with ground state and excited state) and a single mode of a quantized electromagnetic field (e.g., photons in a cavity). The Anti-Jaynes-Cummings (AJC) model is a variant where the interaction terms are reversed. The link between Chiral Dirac oscillators and JC, AJC Hamiltonians, rooted in shared two-level and bosonic structures, enables quantum optical systems (like trapped ions or cavities) to simulate Dirac oscillator physics, including chirality and quantum phase transitions.
\par
\subsection{\centering{Mapping onto a Simultaneous JC-AJC Hamiltonian}}
Now we introduce the circular annihilation-creation operators ($\tilde{a}_r \> \text{and} \> a_l$) defined in \cite{chiral, PhysRevA.76.041801}. Using the axial gauge ($A_0 =0$) the magnetic field in the Dirac oscillator is defined as $\mathbf{B} = B\mathbf{e}_z$ with the vector potential $\mathbf{A} := \frac{B}{2}\left[-y,x,0\right]$. We describe the dynamics by two frequencies, the Dirac oscillator frequency $\omega$ and the cyclotron frequency $\omega_c = eB/m$. Hence, the annihilation-creation operators are:
\[
a_i = \frac{1}{\sqrt{2}}\left( \frac{1}{\triangle}r_i - \iota \frac{\triangle}{\hbar}p_i\right), \tilde{a}_i = \frac{1}{\sqrt{2}}\left( \frac{1}{\tilde{\triangle}}r_i - \iota \frac{\tilde{\triangle}}{\hbar}p_i\right)
\]
where $\triangle := \sqrt{\hbar/m\omega}$, $\tilde{\triangle}:= \sqrt{\hbar/m\tilde{\omega}}$ and $\tilde{\omega} = w_c/2$. The basic commutation relations are,
\begin{equation}
    \begin{split}
        [a_i,\tilde{a}_j] = \frac{1}{2}\left[\frac{\triangle}{\tilde{\triangle}} -\frac{\tilde{\triangle}}{\triangle}\right]\delta_{ij},\\
        [a_i,\tilde{a}_j^{\dagger}] = \frac{1}{2}\left[\frac{\triangle}{\tilde{\triangle}} +\frac{\tilde{\triangle}}{\triangle}\right]\delta_{ij}
    \end{split}
\end{equation}
We have introduced $i = x,y$ to account for two possible directions for equations of motion. So the circular annihilation-creation operators for each frequency are:
\begin{equation}
    \tilde{a}_r = \frac{1}{\sqrt{2}}(\tilde{a}_x - \iota \tilde{a}_y), \> \> \> \>
    \tilde{a}_l = \frac{1}{\sqrt{2}}(\tilde{a}_x + \iota \tilde{a}_y), \> \> \> \>
    a_r = \frac{1}{\sqrt{2}}(a_x - \iota a_y), \> \> \> \>
    a_l = \frac{1}{\sqrt{2}}(a_x + \iota a_y)
\end{equation}
This leads to a bi-chromatic Hamiltonian:
\begin{equation}
    H = \begin{bmatrix}
        1 & \iota \sqrt{2 \xi}a^\dagger_l - \iota \sqrt{2 \tilde{\xi}}\tilde{a_r} \\
        -\iota \sqrt{2 \xi}a_l + \iota \sqrt{2 \tilde{\xi}}\tilde{a_r}^\dagger & 1
    \end{bmatrix}
    \label{2}
\end{equation}
Note that $\tilde{\xi} = \hbar \tilde{\omega}/ mc^2$ and $\xi = \hbar\omega/mc^2$. 
The above Hamiltonian in \eqref{2} can be mapped onto a simultaneous Jaynes Cummings and Anti-Jaynes-Cummings Hamiltonian\cite{JC}. Next we enlist the final results of this mapping and the associated energy spectrum and ground states (next section) obtained in \cite{chiral,PhysRevA.76.041801} to proceed with the discussion of entropy in the third section.
\begin{equation}
    H = \delta\sigma_z - H^{\circlearrowright}_{JC}(g_r) + H^{\circlearrowleft}_{AJC}(g_l)
    \label{h}
\end{equation}
where $\delta = mc^2$ stands for the detuning parameter proportional to the rest mass energy, $H^{\circlearrowright}_{JC}(g_r)$ represents a right-handed Jaynes-Cummings Hamiltonian
\[
H^{\circlearrowright}_{JC}(g_r) = g_r\sigma^+\tilde{a}_r + g_r^*\sigma^-\tilde{a}_r^{\dagger}
\]
with $g_r := \iota mc^2\sqrt{2\tilde{\xi}}$ as the interaction coupling strength. Analogously, the term $H^{\circlearrowleft}_{AJC}(g_l)$ stands for a left-handed Anti-Jaynes-Cummings interaction
\[
H^{\circlearrowleft}_{AJC}(g_l) = g_l\sigma^+a_l^{\dagger} + g_l^*\sigma^-a_l
\]
with a similar coupling strength $g_l := \iota mc^2\sqrt{2{\xi}}$\\

\subsection{\centering{Exact Solution: Energy Spectrum and Eigenstates}}
In this section, we summarize the results of a unitary transformation that converts the bi-chromatic Hamiltonian in \eqref{h} into a monochromatic JC(AJC) term that includes a bosonic degree of freedom with a certain chirality that depends on external parameters $\xi>\tilde{\xi} (\xi<\tilde{\xi})$ \cite{chiral, PhysRevA.76.041801}.

\subsubsection{Left-Handed Regime $\tilde{\xi}< \xi$}
Under the above conditions, the Hamiltonian given in \eqref{h} can be mapped onto a single-mode anti-Jaynes-Cummings Hamiltonian by using the unitary transformation. The above condition also implies that the magnetic field coupling will be dominating, hinting towards a left-handed chirality. 
\begin{equation}
    U_{\alpha} := e^{\alpha(a_l\tilde{a}_r- \tilde{a}_r^{\dagger}a_l^{\dagger})}
    \label{5}
\end{equation}
where the real parameter $\alpha$ depends on the relative strength of the oscillator and magnetic couplings:
\begin{equation}
    \alpha := \frac{1}{\lambda}\text{arctanh}\left(\frac{\lambda\sqrt{\tilde{\omega}}}{\sqrt{\omega} - \mu \sqrt{\tilde{\omega}}}\right) 
    \label{6}
\end{equation}
with $\mu := (\triangle/\tilde{\triangle} - \tilde{\triangle}/\triangle)/2$ and $\lambda := \sqrt{\mu^2 +1}$. The single-mode AJC Hamiltonian will then be $H^{\circlearrowleft}_{AJC}(g_l'):=  U_\alpha HU^\dagger_\alpha=\delta \sigma_z+g'_l \sigma^+a^\dagger_l+(g'_l)^*\sigma^-a_l$ The energy eigenvalues given for this Hamiltonian in \cite{chiral} are: 
\begin{equation}
    E=\pm E_{n_l} = \pm mc^2 \sqrt{1 + 2*\zeta_l(\xi,\tilde{\xi})(n_l +1)}
    \label{7}
\end{equation}
where $\zeta_l(\xi,\tilde{\xi}) := \xi - \tilde{\xi} -2 \mu(\xi\tilde{\xi})^{1/2}$ is related to the initial relevant parameters and $n_l=0,1,...$ represents the number of left-handed quanta. $g'_l=\iota m c^2\sqrt{2\zeta_l}$ is the modified coupling strength. 
After writing the transformation in a monochromatic fashion $U_\alpha = ^{-(\tilde{\mu}\alpha/2)(-a_ra_l+a^\dagger_r a^\dagger_l)}$ where $\tilde{\mu}=(\triangle/\tilde{\triangle}+\tilde{\triangle}/\triangle)/2$. This transformation can be immediately related to a two-mode squeezing operator in the context of quantum optics with squeezing parameter $z:= -\alpha \tilde{\mu}/2 \in \mathbf{R}$ \cite{Barnett:85}. The action of such a squeezing operator \eqref{5} over left-handed chiral Fock states gives rise to SU(1,1) coherent states $|z,n_l\rangle:= U^{\dagger}_{\alpha}|n_l\rangle|vac\rangle_r$ \cite{chiral,Perelomov1972}. 
\begin{equation}
    |z,n_l\rangle = \cosh^{-(n_l + 1)}|z| \sum^{\infty}_{m=0}\sqrt{\frac{(m+n_l)!}{n_l! m!}}(-1)^m \tanh^m|z||m+n_l,m\rangle
    \label{8}
\end{equation}
For the fermionic ground state 
\begin{equation}
    E_g = mc^2
    \label{9}
\end{equation}

\begin{equation}
    |g\rangle = |z,0\rangle|\chi_{\uparrow}\rangle = \frac{1}{\cosh|z|}\sum^{\infty}_{m=0}(-1)^m \tanh^m|z||m,m\rangle|\chi_{\uparrow}\rangle
    \label{9}
\end{equation}
and is interpreted as a spin-up squeezed vacuum state, where the squeezing parameter $z:= -\alpha \tilde{\mu}/2$ depends on the relative coupling strengths $\xi,\tilde{\xi}$.

\subsubsection{Right-Handed Regime $\tilde{\xi}>\xi$}
In this condition, we transform the Hamiltonian into a Jaynes-Cummings Hamiltonian by the unitary transformation. The above limit implies that the oscillatory linear-potential term will be dominating, prescribing the Hamiltonian with a right-handed intrinsic chirality. 
\begin{equation}
    U_{\tilde{\alpha}} := e^{\tilde{\alpha} (a_l\tilde{a}_r - \tilde{a}_l^{\dagger}a_l^{\dagger})}
    \label{10}
\end{equation}
where the parameter $\tilde{\alpha}$ is,
\[
\tilde{\alpha} = \frac{1}{\lambda}\text{arctanh}\left( \frac{\lambda\sqrt{\omega}}{\sqrt{\tilde{\omega}} + \mu\sqrt{\omega}}\right)
\]
In this scenario, the transformed Hamiltonian becomes a single mode Jaynes-Cummings term $H^{\circlearrowright}_{JC}(g_r')=U_{\tilde{\alpha}}HU_{\tilde{\alpha}}^\dagger=\delta\sigma_z + g'_r\sigma^+a_r +(g'_r)^*\sigma^-a_r^\dagger$ with new coupling strength $g'_r:=\iota mc^2\sqrt{2\zeta_r}$ and $\zeta_r(\xi,\tilde{\xi}):=\tilde{\xi}-\xi+2\mu(\tilde{\xi}\xi)^{(1/2)}$. The obtained energy spectrum is then,
\begin{equation}
    \tilde{E} = \pm \tilde{E}_{n_r} = \pm mc^2\sqrt{1 + 2\zeta_r(\xi,\tilde{\xi})(n_r +1)}
    \label{11}
\end{equation}

where $n_r = 0,1,...,$ represents the number of right-handed quanta. The transformation in the single mode frequency domain becomes $U_{\tilde{\alpha}}=e^{(\tilde{\alpha}\tilde{\mu}/2)(-\tilde{\alpha}_r \tilde{\alpha}_l+\tilde{\alpha}_r^\dagger\tilde{\alpha}_l^{\dagger})}$. In this case, the squeezing parameter becomes $\tilde{z} := -\tilde{\alpha}\tilde{\mu}/2 \in \mathcal{R}$ and the SU(1,1) coherent states $|\tilde{z},n_r\rangle := U^{\dagger}_{\tilde{\alpha}} |vac\rangle_l|n_r\rangle $\cite{chiral}.
\begin{equation}
    |\tilde{z},n_r\rangle = \cosh^{-(n_r + 1)}|\tilde{z}| \sum^{\infty}_{\tilde{m}=0}\sqrt{\frac{(\tilde{m}+n_r)!}{n_r! \tilde{m}!}}(-1)^{\tilde{m}} \tanh^{\tilde{m}}|\tilde{z}||\tilde{m}+n_r,\tilde{m}\rangle
    \label{12}
\end{equation}
We also see how the fermionic ground state in this case differs from the left-handed regime one \eqref{9} as:
\begin{equation}
    E_{\tilde{g}} = mc^2\sqrt{1+2\zeta_r}
    \label{14}
\end{equation}
\begin{equation}
    |\tilde{g}\rangle = \tilde{C}_{0+}|\tilde{z},0\rangle|\chi_{\uparrow}\rangle - i\tilde{C}_{0-}|\tilde{z},1\rangle|\chi_{\downarrow}\rangle
    \label{15}
\end{equation}
where $\tilde{C}_{n_r\pm} := [(\tilde{E}_{n_r} \pm mc^2).2\tilde{E}_{n_r}]^{1/2}$, and is interpreted as a spin-up squeezed vacuum state, where the squeezing parameter $\tilde{z}:= -\tilde{\alpha} \tilde{\mu}/2$ depends on the relative coupling strengths $\xi,\tilde{\xi}$.

\subsubsection{Critical Regime $\tilde{\xi} = \xi$}
In this regime, the magnetic field coupling cancels the effect of the Dirac string couple, hence we get a free relativistic fermionic particle \cite{chiral}. The Hamiltonian will be $H^{\text{free}}=c\sum_{j=1}^2 \sigma_j p_j + \sigma_z m c^2$. The energy spectrum is given by:
\begin{equation}
    E^c = \pm E^{c}_{\textbf{p}} = \pm \sqrt{m^2 c^4 + \textbf{p}^2 c^2}
    \label{18}
\end{equation}
where $\textbf{p} = (p_x,p_y)$ stands for two dimensional fermion momentum. We can see that the energy is the usual free particle energy. The eigenstates are described as:
\begin{equation}
    |\pm E^c_{\textbf{p}}\rangle = \sqrt{\frac{mc^2 \pm E^c_{\textbf{p}}}{\pm 2E^c_{\textbf{p}}}}\left( |\chi_{\uparrow}\rangle + \frac{c(p_x + ip_y)}{mc^2 \pm E^c_{\textbf{p}}}|\chi_{\downarrow}\rangle  \right)|\textbf{p}\rangle
\end{equation}
where $\textbf{p} := |p_x,p_y\rangle$ are two-dimensional plane wave solutions. 
\par
Using these results Bermudez et al. in \cite{chiral} showed how the quantum phase transition occurs by plotting the energy spectrum as a function of relative coupling strength, studied the order parameter, the divergence of quantum fluctuations, and the entanglement entropy.  The dependence of the magnetic field provides a control parameter to test the properties of the system such as chirality, squeezing, phonon statistics, and entanglement. 

\section{ENTANGLEMENT AT THE CRITICAL POINT}
Entanglement is one of the key features of a quantum mechanical system. It is a property that manifests a lot of the non-local phenomena in quantum mechanics. In the current system, \cite{chiral} showed that there exists a quantum phase transition, along with angular momentum as the order parameter. Another feature is that in the given left-hand and right-hand regimes, each regime has different ground states. This section is concerned with providing a measure of the entanglement of these states via the entanglement entropy. In the previous work \cite{chiral} the entanglement entropy of these ground states had been calculated using the von Neumann entropy. \\Here, we have tried to use the above-defined SVD entropy to calculate the entanglement entropy. The presence of a left-handed regime ground state and right-handed regime ground state serves as a preselection and postselection state \cite{svd} respectively for the phenomena of quantum phase transition. 
We argue that SVD entropy here can provide a better understanding of the entanglement between the ground states near the critical limit of the quantum phase transition. The SVD entropy can be interpreted as the entanglement entropy across the ground states near the critical point of the quantum phase transition. Since von Neumann entropy is the entropy of the individual states, it does not take into consideration both the left-handed and right-handed regime ground states simultaneously. It quantifies the entanglement between different relativistic degrees of freedom. von Neumann entropy only shows us the behavior of the individual states near the critical point of phase transition meanwhile, SVD entropy shows how the states are getting entangled with the states post-phase transition as the system nears a phase transition. Hence, SVD entropy quantifies the entanglement between different relative degrees of freedom of the different ground states.\\
We begin by defining our Hilbert space. The Hilbert space is a tripartite system, consisting of continuous variables associated with chiral degrees of freedom, and the discrete variables are associated with spin degrees of freedom. To calculate the SVD entropy we would require the transition matrix for the ground states. We define the transition matrix from the given ground states of the left-handed regime $(|g\rangle)$ and the right-handed regime $(|\tilde{g}\rangle)$ \cite{svd}.
\begin{equation}
    \tau^{g|\tilde{g}} = \frac{|g\rangle\langle\tilde{g}|}{\langle g|\tilde{g}\rangle}
\end{equation}
where $|g\rangle$ is the pre-selected state and $|\tilde{g}\rangle$ is the post-selected state.
In order to calculate this matrix. We are considering the ground states.\\
For the \underline{left-handed regime $(\tilde{\xi} < \xi)$} (10,12):
\[
|g\rangle = |z,0\rangle|\chi_{\uparrow}\rangle
\]
and,
\[
|z,n_l\rangle = U^{\dagger}_{\alpha}|n_l\rangle|vac_r\rangle.
\]
So,
\[
|z,0\rangle|\chi_{\downarrow}\rangle = \frac{1}{\cosh|z|}\sum^{\infty}_{m=0}(-1)^m \tanh^{m}|z||m,m\rangle|\chi_{\uparrow}\rangle
\]
For the \underline{right-handed regime $(\tilde{\xi} > \xi)$} (15,17):
\[
|\tilde{g}\rangle = \tilde{C}_{0+}|\tilde{z},0\rangle|\chi_{\uparrow}\rangle - i\tilde{C}_{0-}|\tilde{z},1\rangle|\chi_{\downarrow}\rangle
\]
and,
\[
|\tilde{z},n_r\rangle = U^{\dagger}_{\tilde{\alpha}}|vac_l\rangle|n_r\rangle
\]
\[
|\tilde{z},0\rangle = \frac{1}{\cosh|\tilde{z}|}\sum^{\infty}_{\tilde{m}=0}(-1)^{\tilde{m}} \tanh^{m}|\tilde{z}||\tilde{m},\tilde{m}\rangle
\]
\\
Now,
\begin{equation}
   |g\rangle\langle\tilde{g}| = \tilde{C}^{*}_{0+}|z,0\rangle\langle\tilde{z},0||\chi_{\uparrow}\rangle\langle\chi_{\uparrow}| +\iota\tilde{C}^{*}_{0-}|z,0\rangle\langle\tilde{z},1||\chi_{\uparrow}\rangle\langle\chi_{\downarrow}| 
\end{equation}

and,
\begin{equation}
   \langle g|\tilde{g}\rangle = \frac{\tilde{C}_{0+}}{\cosh|z| \cosh|\tilde{z}|}\sum^{\infty}_{m=0}\tanh^{m}|z| \tanh^{m}|\tilde{z}|. 
\end{equation}

Taking the partial trace of $\tau^{g|\tilde{g}}$ with respect to two out of three indices at a time $l,r,s$ to get three reduced transition matrices.
\begin{equation}
    \begin{split}
        \tau^{g|\tilde{g}}_{l} &= Tr_s(Tr_r \tau^{g|\tilde{g}})\\
        \tau^{g|\tilde{g}}_{l} &= \frac{\frac{\tilde{C}_{0+}}{\cosh|z| \cosh|\tilde{z}|} \sum^{\infty}_{n_l=0} \tanh^{n_l}|z| \tanh^{n_l}|\tilde{z}||n_l\rangle\langle n_l|}{\langle g|\tilde{g}\rangle}    
    \end{split} 
\end{equation}

\begin{equation}
    \begin{split}
        \tau^{g|\tilde{g}}_{r} &= Tr_s(Tr_r \tau^{g|\tilde{g}})\\
        \tau^{g|\tilde{g}}_{r} &= \frac{\frac{\tilde{C}_{0+}}{\cosh|\tilde{z}| \cosh|\tilde{z}|} \sum^{\infty}_{n_r=0} \tanh^{n_r}|z| \tanh^{n_r}|\tilde{z}||n_r\rangle\langle n_r|}{\langle g|\tilde{g}\rangle}        
    \end{split}
\end{equation}

\begin{equation}
    \begin{split}
        \tau^{g|\tilde{g}}_{s} &= Tr_l(Tr_r \tau^{g|\tilde{g}})\\
        \tau^{g|\tilde{g}}_{s} &= |\chi_{\uparrow}\rangle\langle\chi_{\uparrow}|
    \end{split}
\end{equation}

We note that the reduced transition matrix for right and left degrees of freedom is effectively the same. Also, the reduced transition matrix for spinorial degrees of freedom is a pure state hence, its entanglement entropy would be zero ($S_{SVD}^{s} = 0$), this means that there is no spin-orbit entanglement between the discrete and continuous degrees of freedom of both ground states. This value of zero coincides with the left-handed regime von Neumann entanglement entropy for $s$ degree of freedom given in \cite{chiral}. \\
The density matrices for the reduced transition matrices according to the scheme given in \cite{svd} will be:

\begin{equation}
    \begin{split}
        \rho^{g|\tilde{g}}_r &= (1- \tanh|z| \tanh|\tilde{z}|)\sum^{\infty}_{n_r=0} (\tanh^{n_r}|z| \tanh^{n_r}|\tilde{z}||n_r\rangle\langle n_r|)\\
        \rho^{g|\tilde{g}}_l &= (1- \tanh|z| \tanh|\tilde{z}|)\sum^{\infty}_{n_l=0} (\tanh^{n_l}|z| \tanh^{n_l}|\tilde{z}||n_l\rangle\langle n_l|) 
    \end{split}
\end{equation}

Now we calculate the SVD entanglement entropy \cite{svd}
\begin{equation}
    \begin{split}
        S_{SVD}^{\text{l or r}} &= - Tr(\rho^{g|\tilde{g}}_{l \> or \> r} \> \log \rho^{g|\tilde{g}}_{l \> or \> r})\\
        S_{SVD}^{\text{l or r}} &= \frac{\tanh|z| \tanh|\tilde{z}|}{\tanh|z| \tanh|\tilde{z}| -1}\log(\tanh|z| \tanh|\tilde{z}|) - \log(1-\tanh|z| \tanh|\tilde{z}|)
    \end{split}
\end{equation}
when $|z| \approx |\tilde{z}|$,

\newcommand{\cosech}{\mathop{\mathrm{cosech^2}}}

\begin{equation}
    S_{SVD}^{l \> or \> r} = \sinh^2|z| \log (1 + \cosech |z|) + \log(\cosh^2|z|)
\end{equation}
which is exactly equal to the von Neumann entropy of the left-handed regime given in \cite{chiral}.\par
\begin{figure}
    \centering
    \includegraphics[width=0.5\linewidth]{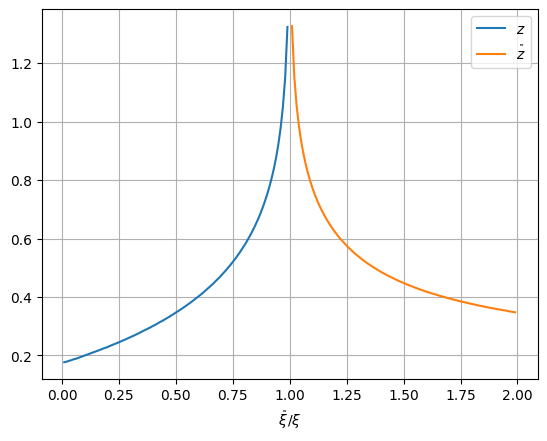}
    \caption{$z$ and $\tilde{z}$ are plotted as a function of the coupling strength ratio. Note that their value diverges at the critical region and the graphs are not mirror images of each other, there is an asymmetry. This asymmetry seems to be an artifact of the model.}
    \label{figure-1}
\end{figure}

Let us focus on the domain of SVD entropy that is $|z|$ and $|\tilde{z}|$. The values for $z$ only exist in one half of the domain (the domain is the coupling strength ratio $\tilde{\xi}/\xi$) and the values for $\tilde{z}$ only exist in the other half of the domain (Figure \ref{figure-1}). This shows that the SVD entropy as a function of both $z$ and $\tilde{z}$ is only valid for points close to the critical point (where $|z| \approx |\tilde{z}|$ and $\tilde{\xi}/\xi \approx 1$) when it is plotted against the coupling strength ratio. 

\begin{figure}
    \centering
    \includegraphics[width=0.5\linewidth]{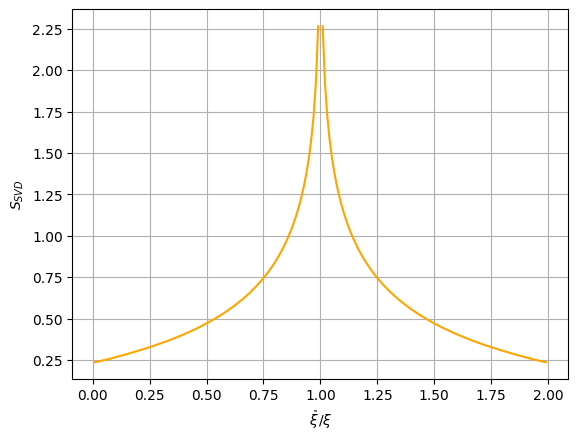}
    \caption{SVD entropy for $l$ and $r$ chiral degrees of freedom is plotted as a function of the coupling strength ratio. The entanglement entropy of the chiral degrees of freedom with the rest of the system diverges at the critical region.}
    \label{figure-2}
\end{figure}
\begin{figure}
    \centering
    \includegraphics[width=0.5\linewidth]{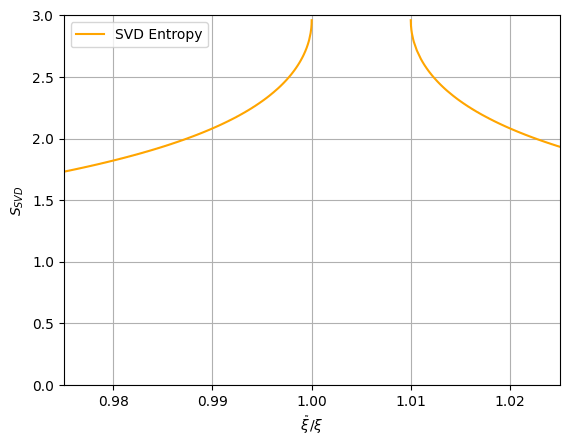}
    \caption{This figure shows more closely the graph for SVD entropy. Near the critical point, the values for SVD entropy become more valid. The point where the graph cuts off is where the value is diverging. Note that the asymmetry present in Figure 1 is also manifest here}
    \label{figure-3}
\end{figure}
\begin{figure}
    \centering
    \includegraphics[width=0.5\linewidth]{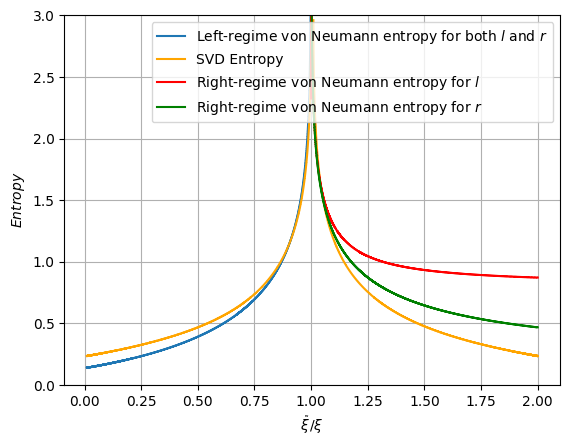}
    \caption{Comparison between SVD entropy and von Neumann entropy for different chiral degrees of freedom. For the left-handed regime, the von Neumann entropy for $l$ and $r$ degrees of freedom is equal \cite{chiral}. Note how the values of left-regime von Neumann and the SVD entropy have the same behavior near the critical point.}
    \label{figure-4}
\end{figure}
\begin{figure}
    \centering
    \includegraphics[width=0.5\linewidth]{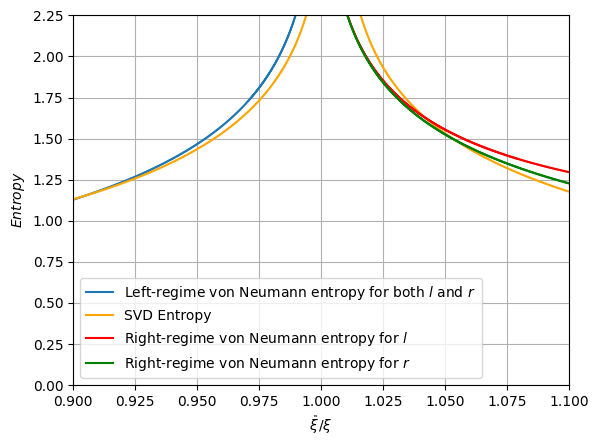}
    \caption{This figure shows more closely the comparison of behaviors between both the definitions of entropies.}
    \label{figure-5}
\end{figure}

The SVD entanglement entropy is plotted (Figure \ref{figure-2}) by considering the values of $z$ and $\tilde{z}$ outside their domain as we have assumed $|z| \approx |\tilde{z}|$. The most accurate representation of SVD entanglement entropy is near the critical point (Figure \ref{figure-3}). Since at the critical point, we can assume that for some value of the coupling strength ratio near the critical point we have approximately equal values of $|z|$ and $|\tilde{z}|$. Numerically, we have taken the values of $\tilde{z}$ starting from the critical point and used those together with the values of $z$(starting from near the origin) to plot the graph for the left-hand side of Figure \ref{figure-2} and similarly, we have taken values from near the origin for $z$ along with values for $\tilde{z}$ (starting from the critical point) to plot the right-hand side of Figure \ref{figure-2}. \\
As shown above for the $|z| \approx |\tilde{z}|$ case we see that the value for the SVD entanglement entropy matches the von Neumann entropy given in \cite{chiral}. This becomes apparent near the critical point of the plot Figure \ref{figure-5}. Figure \ref{figure-4} and Figure \ref{figure-5} is the plot for comparison between von Neumann and SVD entanglement entropy. Note that the value for SVD entanglement entropy is different in the right-handed regime, as compared to the results for von Neumann one. This is an important point of distinction between both the entropies. We also see that SVD entanglement entropy gives us one value for entropy for both $l$ and $r$ degrees of freedom and also for the left-handed and right-handed regime, which arguably makes it a better parameter to study quantum phase transition.

\section{CONCLUSION AND DISCUSSION}
We examined a relativistic spin-1/2 Dirac oscillator in a constant magnetic field. The relativistic Hamiltonian can be mapped to Jaynes-Cummings and anti–Jaynes-Cummings terms \cite{chiral}, describing interactions between the spinor and chirally intrinsic bosons. These models are crucial in several areas of theoretical and experimental research, and reveal the interplay between opposite chiralities.
A. Bermudez et al. in \cite{chiral} showed us two distinct phases under weak and strong magnetic fields, each with opposite chirality. In the intermediate regime, a complex interaction between chiralities leads to a quantum phase transition. In this work we discuss the entanglement entropy of the hybrid ground states obtained from this system across the phase transition in more detail using the SVD entanglement entropy.
\par
The SVD entanglement entropy is a generalization of the von Neumann entropy. It uses a preselection and a post-selection state to calculate the entanglement entropy of the system. Arthur J. Parzygnat et al. in \cite{svd} extended the von Neumann entropy concept from density matrices to general square matrices, defining it as SVD entropy. They showed that SVD entropy possesses similar properties to von Neumann entropy, including additivity, invariance under two-sided independent unitary transformations, and a modified form of concavity. In addition to the examples provided in \cite{svd} we provided a general example (or proof) for SVD entanglement entropy of Bell states which turns out to be $\log2$ similar to the von Neumann case.
\par
The entanglement properties that were brought forth from the SVD definition, turned out to show reasonably similar behavior as the von Neumann case, the values diverge at the critical point which is indicative of the signature of a quantum phase transition. Hence, both definitions are quantitatively similar at times but qualitative differences occur. Further applications of SVD entanglement entropy can be thought about in the field of quantum information and quantum computation theory. This system can also be exploited using the time-like entanglement entropy \cite{narayan2303notes,doi2023pseudoentropy}, where the quantum phase transition can be studied using the time-like generalization offering new physical insight and proposing new phenomena: quantum phase transistion in the time domain.
\par
Entanglement is seen as a measure of complexity from an information theory point of view \cite{Nielsen_Chuang_2010}, for the entanglement entropy to diverge at the critical point, it means that the complexity and the quantum correlations of the degrees of freedom are divergent due to entanglement between the degrees of freedom of left-handed and right-handed regime ground state. This sort of behavior intuitively makes sense in a classical system exhibiting phase transition, as at the critical point the system undergoes a transition hence the information at the critical point of the system is in-deterministic because of its transitory nature \cite{landau2005guide, pethick2008bose, stanley1987introduction, sachdev2011quantum}. For a quantum system, it is explicitly shown by our work and \cite{chiral} how the phase transition happens using entropic calculations, through first principle rather than statistical arguments. 

\section{ACKNOWLEDGEMENT}
One of the authors (YS) has been supported by the Summer Research Program-2024 by S.N. Bose National Centre for Basic Sciences, JD Block, Sector III, Salt Lake, Kolkata 7000106, India.

\printbibliography
\end{document}